\documentclass[twocolumn,showpacs,preprintnumbers,amsmath,amssymb]{revtex4}
\usepackage{graphicx}% Include figure files
\usepackage{dcolumn}% Align table columns on decimal point
\usepackage{bm}% bold math
\usepackage{xspace}
\newcommand{\dist}{\eta_i}
\newcommand{\vol}{\AA$^3$}
\newcommand{\vect}[1]{\mathbf{#1}}
\newcommand{\e}{{\it e}}

\newcommand{\dowidely}{\renewcommand{\baselinestretch}{1.3}\selectfont}
\newcommand{\donarrowly}{\renewcommand{\baselinestretch}{1.0}\selectfont}
\begin{document}

\title{Two-band second moment model and an interatomic potential for caesium}

\author{Graeme J. Ackland and Stewart K. Reed}
\affiliation{School of Physics, \\ The University of Edinburgh, \\
James Clerk Maxwell Building, The King's Buildings,\\
Mayfield Road, Edinburgh EH9 3JZ, UK}
\date{\today}

\begin{abstract}
A semi-empirical formalism is presented for deriving interatomic 
potentials for materials such as caesium or cerium which exhibit volume 
collapse phase transitions.  It is based on the Finnis-Sinclair second moment
tight binding approach, but incorporates two independent bands on 
each atom.  The potential is cast in a form suitable for large-scale 
molecular dynamics, the computational cost being the evaluation of short 
ranged pair potentials.  Parameters for a model potential for caesium 
are derived and tested.
\end{abstract}

\pacs{34.20.Cf , 61.72.-y, 64.60.-i }
\maketitle

\section{Introduction}
Semiempirical models for metallic binding have had a long and successful
history in computer modelling.  The most significant development came in
the mid eighties with the implementation of  `embedded atom'
potentials \cite{db}  based loosely on density functional theory \cite{HKS}
and `N-body' potentials \cite{FS} based on the tight binding second moment
approximation \cite{duc}.  

Although the rationale for these potentials
suggests applicability to free-electron and transition metal systems
respectively,  the implied functional form is similar in each case.
A large number of successful parameterisations of the
functional forms suggested by these methods have been used, and they
have established themselves as the standard method for modelling metallic
systems.

In the second moment approximation to tight binding, the cohesive energy
is proportional to the square root of the bandwidth, which can 
be approximated as a sum of pairwise potentials representing squared 
hopping integrals.  Assuming atomic charge neutrality, this argument 
can be extended to all band occupancies and shapes \cite{avf}.   For simplicity,
consider a rectangular $d$-band of full width W centred on the atomic 
energy level $E_0$.  The  energy for this band relative to the free atom
(the {\it bond energy}) is given by:
\begin{equation}  U_{bond}  = \int_{-W/2}^{E_f=\left(\frac{n}{N}-\frac{1}{2}\right)W }  E\frac{N}{W}\, dE = 
 \frac{W}{2N}n\left(n-N\right)
\end{equation}
where $n$ is the occupation of the band and $N$ the capacity 
(for $d$-bands N=10, for $s$-bands N=2).  The band width
is defined as the square root of the second moment of the density 
of states.  If we project the density of states onto each atom, 
and assume that each atom is charge neutral then the energy becomes:
\begin{equation}
U_{bond} = \sum_i \int_{-W_i/2}^{E_f=( \frac{n}{N}-\frac{1}{2})W_i }  E\frac{N}{W_i} \,dE = 
 - \sum_i\frac{W_i}{2N}n(N-n) \label{eqn:ui} 
\end{equation}
and the second moment of the density of states can be 
calculated as the sum of the squares of the hopping integrals to 
nearest neighbours.  
This latter operation can be written as a sum 
of pair potentials, 
\begin{equation} 
	W_i = \sqrt{\sum_j\phi(r_{ij})}  \label{eqn:phisum}
\end{equation}
The enormous success of Finnis-Sinclair
(and embedded atom-type) potentials arises from their extreme 
computational efficiency: essentially they are no more 
computationally expensive than a conventional pair potential.
This allows them to be applied to extremely large scale simulations, 
addressing complex geometries and phenomena which are intractability with 
more accurate quantum mechanical models.

The computational simplicity follows from the formal division of the 
energy into a sum of energies per atom, which can in turn be evaluated 
locally.
Here we follow a similar philosophy, seeking to incorporate 
as much of the relevant physics as possible, without increasing 
computational cost.

The alkali and alkaline earth metals appear at first glance to be 
close packed metals, forming fcc, hcp or bcc structures at ambient 
pressures.  However, compared with transition metals they are 
easily compressible, and at high pressures adopt more complex 
``open'' structures (with smaller interatomic distances).  
The simple picture of the physics here is of a transfer of electrons 
from an $s$ to a $d$ band \cite{takemura4,abd,macmahan}, the $d$-band being more compact but higher 
in energy.  Hence, at the price of increasing their {\it energy, U}, atoms 
can reduce their {\it volume, V}, and since the stable structure at 
0K is determined by minimum {\it enthalpy, H=U+PV} at high pressure 
this transfer becomes energetically favourable. The net result  is a
metal-metal phase transformation characterised by a large reduction
in volume (and often also in conductivity), the crystal structure itself 
is not the primary order parameter, and in some cases the transition may 
even be isostructural.

The mechanism of the phase transition is unknown.  Although it is obvious 
that an isostructural phase transition is accompanied by an instability 
of the bulk modulus,  the onset of this instability may occur after that 
in the shear modulus.  Thus the mechanism may involve 
shearing rather than isostructural collapse, particularly if a continuous 
interface between the two phases exists, as in a shockwave \cite{shock}.

\section{The Two Band Model}

\subsection{Energies}

%:* Two band model

The Finnis Sinclair formalism 
has been successfully used to study a large variety of
systems. However, we are interested in systems in which electrons
change from one type of orbit to another and in particular, from an
$s$-type orbital to a $d$-type orbital as the sample is pressurised.

\begin{figure}[htb]
%  Energy volume
\centerline{
\includegraphics{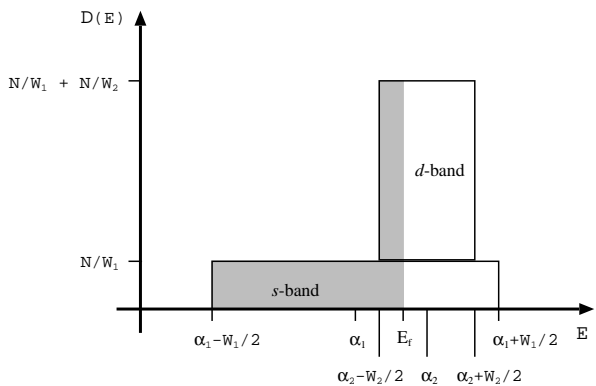}}
        \caption{\label{fig:TwoBands} Schematic picture of DOS in rectangular
 two band model.  Shaded region shows those energy states actually occupied.
}
\end{figure}

Consider therefore two rectangular bands of widths $W_1$ and $W_2$ as
shown in figure~\ref{fig:TwoBands} 
with widths evaluated using equation~\ref{eqn:phisum}.  The bond energy of an atom may be written as the sum of the bond energies of the two bands on that atom
as in equation~\ref{eqn:ui}, and a third term giving the
energy of promotion from band 1 to band 2 (see equation~\ref{eqn:E_promo}):
\begin{equation}
        U_{bond} = \sum_i \frac{W_{i1}}{2 N_{1}}n_{i1}(n_{i1}-N_{1}) +  
        \frac{W_{i2}}{2 N_{2}}n_{i2}(n_{i2}-N_2) + E_{prom} 
        \label{eqn:TwoBondEnergy}
\end{equation}
where $N_1$ and $N_2$ are the capacities of the bands and $n_{i1}$ and 
$n_{i2}$ are the numbers of electrons in each band on the $i^{\text{th}}$ atom.

For an element, enforcing charge neutrality, 
the sum of the numbers of electrons is necessarily
constrained to be equal to the total number of electrons on an atom,
\begin{equation} n_{i1}+n_{i2}=T\,. \label{eqn:tot} \end{equation}
The difference between the energies of the band centres
$\alpha_1$ and $\alpha_2$ is assumed to be fixed. The values of
$\alpha$ correspond to the appropriate energy levels in the isolated
atom. Thus, $\alpha_2 - \alpha_1$ is the excitation energy from one
level to another. For alkali and alkaline earth metals, the free atom 
occupies only $s$-orbitals, so the promotion energy term is therefore simply
\begin{equation}
\label{eqn:E_promo}
        E_{prom} = n_2 (\alpha_2 - \alpha_1) = n_2 E_0 \:,
\end{equation}
where $E_0 = \alpha_2 - \alpha_1$.

Thus the band energy can be written as a function of $n_{i1}$,
$n_{i2}$ and the bandwidths (evaluated at each atom as a sum of 
pair potentials,  within the second moment approximation).  
Defining: 
\begin{equation}\label{eqn:distdef} \dist  = n_{i1}-n_{i2}\end{equation}
 and using equation~\ref{eqn:tot}
we can write:

\begin{widetext}
\begin{equation}
\label{eqn:TwoBandCoh}
        U_{bond} =  \sum_i -\frac{\dist}{4}\left(W_{i1} - W_{i2}\right) 
        - \frac{T}{4}\left(W_{i1} + W_{i2}\right)
+ \frac{\dist^2 + T^2}{8}\left(\frac{W_{i1}}{N_1}+\frac{W_{i2}}{N_2}\right )
+ \frac{\dist T}{4}\left(\frac{W_{i1}}{N_1} - \frac{W_{i2}}{N_2} \right) 
        + \frac{T - \dist}{2}E_0     \:,
\end{equation}
\end{widetext}
Although this expression looks unwieldy, it is in fact computationally
efficient, requiring for its evaluation only two sums of pair potentials for 
$W$ (see equation~\ref{eqn:phisum}) and a minimisation {\it at each site
 independently} with respect to $\dist$.  Since they are local variables, it is
 possible to write closed form expressions for $\dist$ which minimise the 
total energy (see equation \ref{eqn:n_options}).

In addition to the bonding term, the Finnis-Sinclair form includes a 
pairwise repulsion between the ions, which is due primarily to the 
screened ionic charge and orthogonalisation of the valence electrons.  In the present case, 
this pair potential should be a function of $\dist$.  In keeping with 
maintaining locality of the energy we write the pairwise contribution to the energy 
%\begin{eqnarray}U_{pair} & = & \sum_i n_{i1} U_{pair_{i1}} + n_{i2} U_{pair_{i2}} \nonumber \\        & = & \sum_{ij}V(r_{ij},\dist) + V(r_{ij},\eta_j) \end{eqnarray}
 in the intuitive form, as the sum of two terms,  one from each ``band'', proportional to the number of electrons in that band
\begin{equation}V(r_{ij}) =  (n_{i1}+n_{j1}) V_1(r_{ij}) + (n_{i2}+n_{j2}) V_2(r_{ij})\label{eq:ppot}\end{equation}
We rearrange this to give the energy as a sum over atoms
\begin{equation}
U_{pair} 
=
\sum_{i}\left [ n_{i1} \sum_{j\ne i} V_1(r_{ij}) + n_{i2}\sum_{j\ne i}  V_2(r_{ij}) \right ]
\end{equation}

The total energy is now simply
\begin{equation}
U_{tot} =  U_{pair} + U_{bond}  \;.
\end{equation}
This depends on $\dist$, but as described above the $\dist$ take the values
which minimise the energy and one can solve:
\begin{equation}
\label{eqn:dEcohdn0}
        \frac{\partial U_{tot}}{\partial \dist} = 0 
\end{equation}
explicitly for $\eta_{i_0}$ independently at each atom, whence
\begin{widetext}
\begin{equation}
        \label{eqn:mu0}
        \eta_{i_0} = \frac{N_1 N_2}{W_{i1} N_2 + W_{i2} N_1} 
        \left[ W_{i1} - W_{i2} - T\left(\frac{W_{i1}}{N_1} - \frac{W_{i2}}{N_2}\right) 
        + 2E_{0} -2U_{i1,pair} +2U_{i2,pair} \right] \,.
\end{equation}
\end{widetext}

Depending on the number of electrons in the system it may not be possible to realise this. Charge neutrality  requires that $\vert \dist \vert$ cannot be greater than the total number of electrons $T$ per atom.  The fixed capacities of the bands  ($N_1$ and $N_2$) can also prohibit the realisation of $\eta_{i_0}$. It is therefore necessary to limit the values which $\dist$ may have:
\begin{equation}
\dist = 
\label{eqn:n_options}
\begin{cases}
             {\rm min}(T,2N_1-T),  & \text{if $\;\eta_{i_0} > {\rm min}(T,2N_1-T)$}\\
                {\rm max}(-T,T-2N_2), &\text{if $\;\eta_{i_0} < {\rm max}(-T,T-2N_2)$}\\
                \eta_{i_0}, & \text{otherwise}
\end{cases}
\end{equation}
where $\eta_{i_0}$ is given by equation~\ref{eqn:mu0}.  
The expressions for $\dist$ involves only constants and sums of pair 
potentials, and can be evaluated independently at each atom at a similar 
computational cost to a standard many-body type potential.

\subsection{Forces}
Since the energy is variational in $\dist$ within this model, we have
\begin{equation} \label{eq:HF}
\frac{\partial{U}_{tot}}{\partial\dist}=0
\end{equation}
and the force on the $i^{\rm{th}}$ atom is
\begin{eqnarray}\label{eqn:HF}
\vect{f}_{i}& = & - \frac{dU_{tot}}{d{\bf r}_i} \nonumber \\
        & = & -\frac{\partial U_{tot}}{\partial {\bf r}_i}\bigg\vert_{\eta} - 
\frac{\partial U_{tot}}{\partial \eta}\frac{\partial \eta}{d {\bf r}_i} \nonumber \\
        & = & -\frac{\partial U_{tot}}{\partial {\bf r}_i}\bigg\vert_{\eta} 
\end{eqnarray}
Hence the force is simply the derivative of the energy at fixed $\eta$.
Basically, this is the Hellmann-Feynman theorem \cite{p:Feynman} which arises here because
$\eta$ is essentially a single parameter representation of the electronic 
structure. 

This result means that, like the energy, the force can be evaluated by
summing pairwise potentials.  Hence the model is well suited for 
large scale molecular dynamics.

The force derivation is somewhat tedious (see appendix~\ref{appendix-force}), the result
being:
\newlength{\oldcolsep}
\setlength{\oldcolsep}{\arraycolsep}
\setlength{\arraycolsep}{0cm}
\begin{eqnarray}
{\bf f_i} = \sum_j  &&
 \Big[(\tilde{W_{i1}}+\tilde{W_{j1}}) {\bf \phi}_1'(r_{ij}) 
   + (\tilde{W_{i2}}+\tilde{W_{j2}}) {\bf \phi}_2'(r_{ij})
\nonumber\\
 && {} 
- (n_{i1} + n_{j1}){\bf V'}_1(r_{ij}) 
   - (n_{i2} + n_{j2}){\bf V'}_2(r_{ij}) 
 \Big]\hat{\bf r}_{ij} \nonumber\\
\end{eqnarray}
\setlength{\arraycolsep}{\oldcolsep}
with 
\[ \tilde{W_{ib}} = \frac{n_{ib}(N_b-n_{ib})}{4N_bW_{ib}}  \]
and where the numbers of electrons in each band, $n_{ib}$, have been calculated analytically 
using equations \ref{eqn:mu0}, \ref{eqn:n_options} and \ref{eqn:distdef}. 
The contribution to the force from each neighbour acts along the direction 
of the vector between the atoms.

\section{A Model Potential for Caesium}
One application of the model is to simulate the pressure-induced 
isostructural phase transition in materials such as caesium.  Here
the transformation arises from electronic transition from the $s$ to 
the $d$-band.  Caesium adopts the bcc structure (Cs I) at ambient pressure,
transforming under slight additional pressure to fcc (Cs II) and 
under further increase transforming isostructurally to fcc (Cs III) \cite{p:HallMerrillBarnett}
\footnote{There is some recent evidence that the Cs III phase may be a more complex structure \cite{p:MalcomCsIII}.}.
Here we seek to represent the volume collapse Cs II $\rightarrow$ Cs III transition.

\subsection{Parametrisation}
To make a usable potential, the functional forms of $\phi$ and $V$ 
must be chosen.  Although this is somewhat arbitrary, the physical 
picture of hopping integral and screened ion-ion potential suggests
that both should be short ranged, continuous and reasonably smooth.

In the present case, we are interested in the gross features of the
model, so we adopt simple forms fitted to the cohesive energies and
volume of the high and low pressure polytypes of
caesium.  For more complex applications, it may be necessary to 
fit other properties.

Following Finnis and Sinclair, we choose for the hopping integral
\begin{equation}
\phi_b(r_{ij})=
\begin{cases}
C_b (d_b-r_{ij})^3,&\text{if $r_{ij} \leq d_b$}\\
0, &\text{if $r_{ij} > d_b$}
\end{cases}
\end{equation}
and for the pairwise part, following Lennard-Jones
\begin{equation}
\label{eqn:rep_i}
        V_b(r_{ij}) = \sum_j \frac{A_b}{r_{ij}^{12}}
\end{equation}
where $b$ labels the band.

We select the promotion energy $E_0$ to be that required to promote an electron from the $6s$ level into the $5d$ level of an isolated atom~\cite{b:CRCHandbook,calc:Simon}.  The band capacities are $N_s=2$, $N_d=10$
and the total number of electrons per atom is $T=1$. The cutoff radii $d_s$ and $d_d$ are chosen to be between the second and third nearest neighbours and between the first and second nearest neighbours respectively. The remaining four parameters ($A_s$, $A_d$, $C_s$, $C_d$) are available for fitting. 

\emph{Ab initio} calculations using pseudopotentials, plane waves and the
generalized gradient approximation~\cite{xie,SKR:thesis} show that the
energy-volume relations for bcc and fcc caesium are almost degenerate,
so we have fitted the remaining parameters to the atomic volumes of Cs~I 
(115.9 \AA$^3$/atom), Cs~II (67.5 \AA$^3$/atom) and Cs~III
(48.7 \AA$^3$/atom) \cite{b:CRCHandbook} as well as the transition
pressure between phases II and III (4.3 GPa) \cite{b:CRCHandbook} and
the cohesive energy of phase I (-0.704 eV). The cohesive energy is the
sum of the heat of formation and the heat of
vaporisation~\cite{b:CRCHandbook}.

Initial estimates for the parameters were determined using a symbolic mathematics package. The final parameters were arrived at by an iterative process: Least squares,  conjugate gradients and \emph{ab oculo}  minimisation techniques were used to determine the best parameters for particular cutoff radii. The cutoff radii were then adjusted by hand to improve the fit. This process was repeated until the optimum fit was achieved. The final parameters are given in table~\ref{table:TB-Params}.

\begin{table}
\begin{center}
\dowidely
\begin{tabular}{|c|c||c|c|}
\hline
\multicolumn{2}{|c||}{$s$ band}& \multicolumn{2}{c|}{ $d$ band} \\
\hline
\hline $C_s$ & 0.05617 eV$^2$\AA$^{\mathrm{-3}}$  & $C_d$ & 0.1681 eV$^2$\AA$^{\mathrm{-3}}$ \\
\hline $d_s$ & 9.5097 \AA      & $d_d$ & 6.9189 \AA\\
\hline $A_s$ & $2.4017\times 10^7$ eV \AA$^{12}$ & $A_d$ & 3.7668$\times10^6$eV \AA$^{12}$\\
\hline
\hline
\multicolumn{2}{|c||}{E$_0$}& \multicolumn{2}{c|}{ 1.19 eV} \\
\hline
\end{tabular}
\donarrowly
%\parbox[m][1in][c]{1in}{ \centerline{E$_0$ = 1.19}}
\end{center}
\caption{\label{table:TB-Params}Parameters for the two band model, 
obtained by fitting to phases II and III of caesium.}
\end{table}

\begin{figure}[htb]
%  Energy volume
\centerline{
\includegraphics{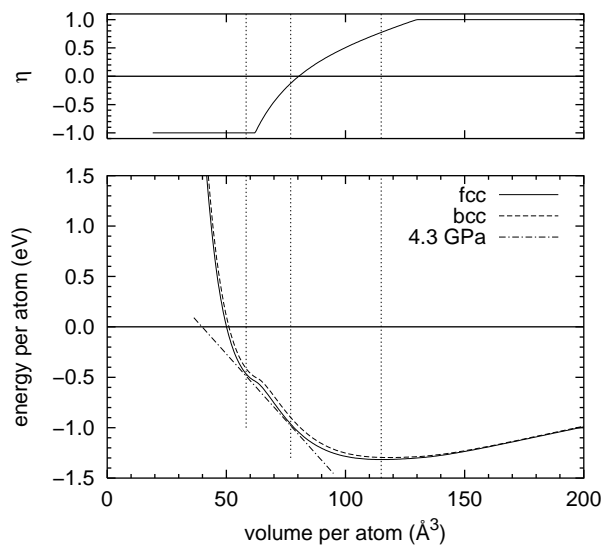}
}
        \caption{\label{fig:TB-EngVol} Top: Variation of $\dist$ with 
compression, showing $s \rightarrow d$ transfer in model caesium ($T=1$). 
Bottom: An energy-volume curve for the two 
band potential. The minimum lies at -1.3163 eV and 115.2 \AA$^3$/atom. 
The gradient of the straight dash-dotted line is the experimental fcc-fcc
transition pressure.  In reality caesium also has a bcc-fcc phase transition at 2.3 GPa, however
first principles calculations show that these two structures are almost 
degenerate in energy at 0K.  The isostructural transition, which involves 
electron localisation and hence is not properly reproduced by standard
electronic structure calculation, occurs at 4.3 GPa.
}
\end{figure}

Figure~\ref{fig:TB-EngVol} shows the energy-volume curves for the fcc
and bcc structures calculated using the model.
 The fcc structure is stable everywhere compared to the bcc structure. 
Experimentally however, Cs~I has bcc structure with an equilibrium volume of 116 \vol.  The present
potential does not have a low pressure bcc phase. However, at ambient pressure 
the bcc and fcc curves are almost degenerate (0.02 eV/atom difference) in agreement with the \emph{ab initio}
 calculations.
The predicted equilibrium volume of the ambient pressure phase agrees well with the experimental 
equilibrium volume of phase I of caesium. Although the predicted
volumes of phases II and III are larger than the experimental values, the predicted transition pressure and volume collapse are in good agreement with experiment.

\subsection{Elasticity}

The elastic moduli are not fitted explicitly, so their behaviour represents a 
sensitive test of the model.  Application of the pressure generalisation of the 
Born stability criteria is rather confused in the present literature 
\cite{yip,bbk,marcus} so we lay this out in appendix~\ref{appendix-elastic}.

Since caesium exhibits an isostructural phase transition the bulk modulus must 
formally become negative at some volume (where the structure is unstable).
It is less clear whether the other Born stability criteria will be violated: 
{\it ab initio} density functional perturbation theory calculations \cite{xie} find an
 instability in the long range acoustic phonons: at zero pressure this is equivalent
 to an instability in $C'=(C_{11}-C_{12})/2$.

\begin{figure}[htb]
%  C_11, C_12, C_44, (C_{11}-C_{12})/2
\centerline{
\includegraphics{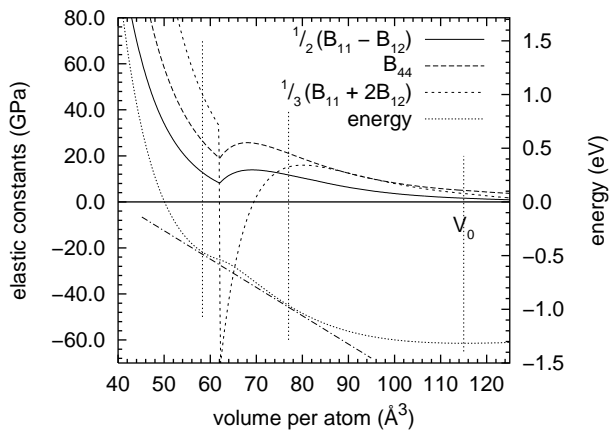}}
        \caption{\label{fig:BulkVol}Graph of elastic stiffness constants 
against volume.
Mechanical instabilities occur for negative values of $B_{44}$, ($B_{11}-B_{12})/2$ and 
$(B_{11}+2B_{12})/3$.  Total energy is shown on the same figure but with respect to the right hand axis. The right most vertical line indicates the volume at ambient pressure. The other two vertical lines 
delineate the volumes in which the structure is unstable with respect to decomposition into coexisting phases whilst the straight dash-dotted line is the transition pressure from phase II to phase III. 
}
\end{figure}

With the present potential the analytic expressions for the 
elastic constants are complicated: there is no simplification 
akin to equation \ref{eq:HF} for the second derivatives.  Consequently, 
we evaluate the elastic constants numerically from finite strain
calculations, see figure~\ref{fig:BulkVol}.  
We find that the volume collapse is announced by a slight softening of the bulk
 modulus which then goes negative in the unstable region. Although the shear 
and tetragonal shear decrease in the unstable region, neither actually goes
 negative.   

\subsection{Defects}

One of the major features of many-body and embedded atom type
potentials is their ability to describe defects such as surfaces,
vacancies and interstitials without the constraints of pair
potentials, for which the undercoordination defect energy is simply
the sum of broken bond energies, less a small amount from relaxation
of atomic positions.  Thus the vacancy formation energy is typically
the same as the cohesive energy, whereas in real materials it is
typically less than half.  The present potential is not fitted to any
defect configuration, so it is of interest to see what it predicts -
moreover the variety of environments associated with defects provides
a good check against pathologies.

The relaxed surface energies for four low-energy surfaces are given in
table~\ref{table:Surf}.  As is usual with many-body potentials the
energy is much lower than would be expected from simple bond
counting. For the (111) surface the energy for each atom in the plane
is increased by about $1/9$ of its cohesive energy, while each atom has 1/4 of its
bonds broken.  There is also transfer of all electrons from $d$-like to $s$-like
states at the free surface.  This also leads to an unusual outwards 
relaxation of the surface atoms.

Likewise the vacancy formation energy and volume
(table~\ref{table:Int}) are rather typical of many-body potential
results.

The interstitial formation energy is especially low (table~\ref{table:Int}),  around 1.8 eV
depending on the orientation and the amount of relaxation allowed locally,
meaning that thermal interstitial formation is possible.  This 
low value is due to the transfer of electrons from $s$- to $d$-band
on the interstitial atom or dumbbell, similar to the high pressure 
behaviour.  Consistent with recent \emph{ab initio} calculations in various elements 
\cite{madden,becquart,SWHan}, the calculated interatomic spacing of the 
dumbbell atoms is much smaller than in the bulk, (around 15\%) and much
 smaller than is typical of standard EAM-type  potentials.  As a consequence 
of this, the associated strain fields are considerably smaller.

For a detailed study of point defects in caesium, it would be
appropriate to reparametrise the potential with point defects
included in the fitting, but the good results obtained here 
without such fitting suggest
that the present model contains the right physics.

\begin{table}
\begin{center}
\dowidely
\begin{tabular}{|c|c||c|c|}
\hline surface & energy meV/\AA$^2$   & relaxation 1\AA & relaxation 2\AA\\
\hline
\hline 111 &  5.938     & -0.042 & 0.000 \\
\hline 001 &  6.54       & -0.046 & -0.007 \\
\hline 011 & 7.195 & 0.024  & -0.015 \\
\hline 211 & 6.952 & -0.122  & -0.045 \\
\hline
\end{tabular}
\donarrowly
\end{center}
\caption{\label{table:Surf}Surface energies, and atomic relaxations of the top two layers in angstroms}
\end{table}

\begin{table}
\begin{center}
\dowidely
\begin{tabular}{|c|c|}
\hline configuration & energy $eV$ \\
\hline
\hline 111 dumbbell &  2.178  \\
\hline 001 dumbbell &  1.776     \\
\hline 011 dumbbell &  1.975 \\
\hline 011 crowdion &  1.975 \\
\hline vacancy & 0.545\\
\hline
\end{tabular}
\donarrowly
\end{center}
\caption{\label{table:Int}Interstitial and vacancy formation energies. The vacancy formation volume is 0.846V$_0$.}
\end{table}

\section{Extrapolation to transition metals}

In principle, the current formalism should be applicable to $d$-band metals.
We do not intend to refit the potentials here, but by applying the 
parameters fitted for caesium with appropriate scaling for ionic charge,
and simply varying the total number 
of electrons we recover the parabolic behaviour of the cohesive energy 
and bulk modulus which characterises the transition metal series.  
Considering that there are no fitting parameters, the agreement 
with experimental results is extraordinarily good.
Moreover, the volume collapse phase transition exists only for N=1 
and N=2 (consistent with experiment).  However, since no information 
about band shape is included, the sequence of crystal structures 
cannot be reproduced, so we consider here only the fcc structures, 
calculating their ``experimental'' values from the experimental density.

While the extrapolated potentials do not represent the optimal 
parametrisation for specific transition metals, the recovery of the 
trends across the group lends weight to the idea that the 
two-band model correctly reproduces the physics of this series.

\begin{figure}[htb] 
\includegraphics{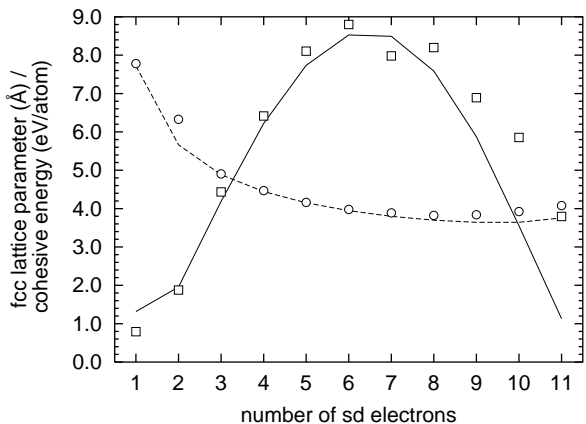}
\caption{\label{fig:trans}
Extrapolation of model to various {6s5d}-band materials.  The
parameters $E_0$, $A_s$, $A_d$, $C_s$, $C_d$, $d_s$ and $d_d$
are taken from the fit to caesium, with the lengths being scaled 
according to the Fermi vector ($Z^{-1/3}$) and the energies by  
($Z^{1/2}$).  The lattice parameters are shown by the
solid line (calculated) and circles (experiment) and the 
cohesive energies by dashed line (calculated) and squares 
(experiment).  Calculations are actually done at integer values of 
$T$ and lines join these points.}
\end{figure}

\section{Conclusions}

We have presented a model to describe $s \rightarrow d$ transfer within the 
framework of the empirical second-moment tight binding model. 
With a very simple parametrisation, our model describes the 
isostructural phase transition and associated $s \rightarrow d$ 
transfer in Cs and allows study of the elastic instabilities which 
occur under pressure and uniaxial stress.  Throughout, our formalism 
is guided by the principle of retaining as much electronic detail as 
possible within the constraint of the computational complexity of 
short-ranged pair potentials

Our potential form involves evaluation of sums of pair potentials, 
and minimisation of the energy with respect to band occupation.  In the 
approximation of local density of states evaluated at each atom, it 
becomes possible to formally write the energy of each atom in terms 
of the local band occupation, a single parameter $\dist$.  This 
variational parameter can be explicitly eliminated from the energy and
 force expressions, meaning that energies and forces can be evaluated 
by summing pair potentials.  

To our knowledge,  the present potential represents the most sophisticated 
energy function which has first derivatives which are analytic pair potentials, 
and the first exploitation of  the Hellman-Feynman theorem in empirical 
potentials to achieve this. It is thus uniquely well suited for exploitation in 
standard molecular dynamics codes.

Extension of the formalism to transition metals is straightforward,
and the caesium parametrisation gives a surprisingly good description 
across the group. While providing impressive proof of concept, this may not
 be of enormous practical use, however, since once there is a significant 
partial occupation of the $d$-band under all conditions the model is similar
 to the standard many-body potentials which are known to do a good job in
 describing transition metals.

The precise stable crystal symmetry depends on the shape of the band 
structure, and is not therefore correctly described in this rectangular 
band model.  Many-body potentials never contain the correct physics to 
describe phase transitions, except in the martensitic case of freezing 
in soft phonons \cite{Pinsook,Pinsook2}.  However, the crystal structure which 
has the lowest energy can be determined by judicious choice of functions
$V$ and $\phi$.  Here the simple use of power law repulsion and cubic
attraction leads to close packed structures.

The two-band formalism can be easily combined with alternate 
parameterisations to get effective band shapes, and could be applied  
to ferromagnetic systems where the two bands represent separate 
spins, and the atomic $E_0$ term is replaced by a term which favours 
maximum spin.

%\newpage
\appendix

\begin{widetext}
%Appendix A: 
\section{Elastic instabilities under pressure - energy definition}
\label{appendix-elastic}
There is a degree of confusion in the literature regarding the definition 
of elastic constants under pressure.  To some extent this arises because of 
various definitions of strain (Lagrangian, Eulerian, volume conserving).  
Here we lay out the definitions in terms of energy.  One important point 
to note is that under pressure the moduli which correspond to long wavelength
phonons are volume conserving, while those corresponding to crystal stability 
are not.  At finite pressure, this means they are different.

At pressure, the important quantity for stability in energetic terms is 
the Gibbs free energy $G$.
\begin{equation}
G = U - TS + PV
\end{equation}
where P is an externally applied hydrostatic pressure. $U-TS$ is the
Helmholtz free energy F. The change in the Gibbs energy due to a
distortion of the crystal is
\begin{equation}
\Delta G = \Delta F + \Delta P.V + P \Delta V \;.
\end{equation}
If the hydrostatic pressure is applied by an external mechanism then
$\Delta P$ is zero. Therefore
\begin{equation}
\Delta G = \Delta F  + P \Delta V \;.
\end{equation}

Any arbitrary change in the unit cell 
be expressed in terms of the strain tensor 
$\stackrel{\leftrightarrow}{\varepsilon}$.
Using Voigt notation, the matrix representation of the strain tensor 
is written as
\begin{equation}
\label{eqn:Voigt-strain}
\stackrel{\leftrightarrow}{\varepsilon}
=\left(\begin{array}{ccc} e_1   &  e_6/2 &  e_5/2 \\e_6/2 &  e_2  
  &  e_4/2 \\e_5/2 &  e_4/2 &  e_3 \end{array}\right)\,.
\end{equation}

If we write the equilibrium cell as a matrix comprising the three lattice vectors
 $\stackrel{\leftrightarrow}{V}_0 = ({\vect{r}_1 \; \vect{r}_2 \; \vect{r}_3 })$, then 
any arbitrary strain give a new unit cell:
\begin{equation}
 \stackrel{\leftrightarrow}{V} =  \stackrel{\leftrightarrow}{V}_0(1+\stackrel{\leftrightarrow}{\varepsilon}) \;.
\end{equation}

If the volume of the original cell is 
$V_0 = |\stackrel{\leftrightarrow}{V_0}|$, 
then the volume of the new cell is
\begin{eqnarray}
V/V_O &=&  |\stackrel{\leftrightarrow}{V}| /  |\stackrel{\leftrightarrow}{V}_0| \\
  &=& 1 + e_1 + e_2 + e_3 + e_1e_2 + e_2e_3 + e_3e_1 \nonumber \\ 
&& - e_4^2/4 - e_5^2/4 -e_6^2/4 \nonumber \\ 
&& + e_1e_2e_3 -e_1e_4^2/4 - e_2e_5^2/4 - e_3e_6^2/4 + e_4e_5e_6/4 \;. \nonumber
\end{eqnarray}

Which can be expressed concisely in standard strain notation 
($i,j,k,l$ representing cartesian directions) as:
\begin{equation}
\frac{\Delta V}{V} = \e_{ii} 
+ \frac{1}{4}\left( 2\delta_{ij}\delta_{kl} - \delta_{ik}\delta_{jl} - \delta_{il
}\delta_{jk}\right) \e_{ij}\e_{kl}
+ \mathcal{O}(\varepsilon^3)
\end{equation}
where $\delta_{ij}$ is the Kroneker delta and the usual implicit sum
convention for tensors applies. The change in the Gibbs free energy may then be written
\begin{equation}
\label{eqn:DGequals}
\Delta G = \Delta F + P\e_{ii} 
+ \frac{PV}{4}\left( 2 \delta_{ij}\delta_{kl} - \delta_{ik}\delta_{jl} - 
\delta_{il}\delta_{jk}\right) \e_{ij}\e_{kl}\;.
\end{equation}

We define $C_{ij}$ as the second order derivatives of the 
Helmholtz free energy with respect to the Voigt strains. 
and $B_{ij}$ as the second order derivatives of the 
Gibbs free energy with respect to the Voigt strains. 
The limiting case of long wavelength phonons is related to 
$C_{ij}$ which can be calculated from the dynamical matrix \cite{sjc}, 
the crystal stability criteria are related to the  $B_{ij}$ \cite{bbk}.
In the standard notation:
\begin{equation}
\label{eqn11}
B_{ijkl} = C_{ijkl}+\frac{P}{2}\left( 2\delta_{ij}\delta_{kl} - \delta_{ik}\delta_{jl} - \delta_{il}\delta_{jk}\right) \;.
\end{equation}
In Voigt notation:
\begin{equation}
B_{ij} = 
\begin{cases}
%     C_{ii} \hspace{1in} \text{if $\; i=j$ and $\; i<4$}\\
     C_{ii} -P/2 \hspace{0.6in} \text{if $\; i=j$ and $\; i>3$}\\
     C_{ij} +P \hspace{0.7in} \text{if $\; i \neq j$ and $\; i,j<4$}\\
     C_{ij} \hspace{1in}  \text{otherwise}\\
\end{cases}
\end{equation}

The Born elastic stability criteria restrict what values various moduli may 
have if the crystal is to be stable~\cite{b:Born}.  In the case of external pressure
 the relevant free energy is $G$ and the relevant moduli are $B_{ij}$.  Due to the
 Voigt symmetry, the elastic constant tensor has in general, only 21 independent
 components. In cubic crystals however this number is further reduced by symmetry
 to 3, namely $B_{11}$, $B_{12}$ and $B_{44}$.  Consequently, there are three 
 common stability criteria which impose lower bounds on the bulk, shear and 
 tetragonal shear  moduli of cubic crystals. These may be written in Voigt notation:
\begin{equation}
\frac{1}{3}(B_{11}+2B_{12})> 0\,,
\quad B_{44} > 0
\quad \mathrm{and}
\quad\frac{1}{2}(B_{11}-B_{12}) >0 \,.
\end{equation}
Whence the generalised Born stability criteria for the elastic constants 
of cubic crystals at pressure become:
\begin{eqnarray}
\frac{1}{3}(C_{11}+2C_{12}+2P) &>& 0\,, \nonumber \\
C_{44}-\frac{p}{2} &>& 0 \,,\nonumber \\
\quad\frac{1}{2}(C_{11}-C_{12}-P) &>&0\,,  \end{eqnarray}
where
\begin{equation}B_{11} = C_{11}\,,\quad B_{12}
 = C_{12}+P\,\quad\mathrm{and}\quad B_{44} = C_{44}-\frac{p}{2}\,. 
\end{equation}

\subsection*{Calculation of bulk, shear and tetragonal shear moduli from energies}
In this paper we evaluate the elastic moduli by applying finite strain
and measuring the change in total energy $U_{tot}$.  The finite strain 
method has the advantage over analytic differentiation of automatically 
compensating for non-isotropic movement of the atoms.  At zero
temperature $U_{tot}$ is the same as the classical Helmholtz free energy
$F$.  At finite pressure, however, the system is not at a minimum of
$U_{tot}$ with respect to strain.

When applying strain to a system under pressure, the 
first order term in the Gibbs free energy vanishes:
\begin{equation}
\Delta G = \e_{ij} \frac{\partial G}{\partial \e_{ij}}
+\frac{1}{2}\e_{ij}\e_{kl}\frac{\partial^2G}{\partial \e_{ij}\partial\e_{kl}}
=\frac{1}{2}\e_{ij}\e_{kl}\frac{\partial^2G}{\partial \e_{ij}\partial\e_{kl}}
\end{equation}
But the first order term for Helmholtz does not.
For a cubic crystal, one can find $(C_{11}+2C_{12})/3$, 
$(C_{11}-C_{12})/2$ and $C_{44}$ directly at any volume by simple
distortions and measurement of $F$.

For bulk modulus we apply a pair of Voigt strains,  with $\e_{1}= \e_{2}=\e_{3}=e$, 
the other elements being zero. Assuming we make a pair of small distortions about 
a volume V,  the difference 
\begin{equation}
G(e) + G(-e) - 2G(0) = 
V \frac{\left(B_{11}+2B_{12}\right)}{3}9e^2 + \mathcal{O}(e^3)\;.
\end{equation}
equivalently, using the Helmholtz free energy
\begin{equation}
\frac{\left( C_{11}+2C_{12}\right)}{3} = 
\frac{(F(e) + F(-e) - 2F(0) )}{9Ve^2}
+ \mathcal{O}(e^3)\;.
\end{equation}

For  $(C_{11}-C_{12})/2$, the relevant distortion, volume conserving to first order, is $e=e_1=-e_2$ and
\begin{equation}
\frac{\left( C_{11}-C_{12}\right)}{2} = 
\frac{(F(e) + F(-e) - 2F(0) )}{4Ve^2}
+ \mathcal{O}(e^3)\;.
\end{equation}

For  $(C_{44}$, the relevant distortion is $e=e_6$ and
\begin{equation}C_{44} =
\frac{( F(e) + F(-e) - 2F(0) )}{Ve^2} + \mathcal{O}(e^3)\;.
\end{equation}

\section{Derivation of the force on an atom}
%Appendix B: 
\label{appendix-force}
\renewcommand{\dist}{\eta}

Since the cohesive energy is variational with respect to the electron distribution 
$\dist$, the derivative of the energy with respect to the electron distribution is zero
and so the force may be written as a partial derivative akin to the Hellmann-Feynman
theorem~\cite{p:Feynman}.  From equation~\ref{eqn:HF}
\begin{equation}
\vect{f}_i = \frac{\partial U_{tot}}{\partial\vect{r}_i}\bigg\vert_{\dist} = -\sum_j\frac{\partial U_j}{\partial\vect{r}_i}\bigg\vert_{\dist_j}\;.
 \end{equation}
where the sum over $j$ includes $j=i$.
From equations \ref{eqn:TwoBondEnergy}, \ref{eqn:E_promo} and \ref{eqn:TwoBandCoh},
the cohesive energy of the $j^{th}$ atom in terms of the number of electrons in each band 
($n_{j1}$ and $n_{j2}$) is
\begin{equation}
%\begin{aligned}
U_j =  \bigg[  \frac{n_{j1}}{2}\bigg(\frac{n_{j1}}{N_1} - 1\bigg) W_{j1}
+\frac{n_{j2}}{2}\bigg(\frac{n_{j2}}{N_2} - 1\bigg) W_{j2} +
  n_{j1} \sum_{k\ne j} V_1(r_{jk}) + n_{j2}\sum_{k\ne j}  V_2(r_{jk}) 
 + n_{j2} E_0\bigg] \,,
%\end{aligned}
\end{equation}
and the force on atom $i$ is thus
\begin{equation} \vect{f}_i = - \sum_j  \bigg[ \frac{n_{j1}}{2}\bigg(\frac{n_{j1}}{N_1} - 1\bigg) \frac{\partial W_{j1}}{\partial \vect{r}_i}
 +\frac{n_{j2}}{2}\bigg(\frac{n_{j2}}{N_2} - 1\bigg) \frac{\partial W_{j2}}{\partial \vect{r}_i}
 +\left ( n_{j1} + n_{i1} \frac{\partial}{\partial \vect{r}_i}  V_1(r_{ij}) \right ) 
+\left ( n_{j2} +n_{i2}  \frac{\partial}{\partial \vect{r}_i}  V_2(r_{ij})\right )  \bigg]\,.
\end{equation}
%From equation~\ref{eqn:width_i}, 
The width of band b on atom $i$ is \( W_{jb} = \sqrt{\sum_{k \neq j} \phi_b(r_{jk})} \) and its derivative is 
\begin{equation}\frac{\partial W_{jb}}{\partial \vect{r}_i} 
= \frac{1}{2W_{jb}}\sum_{k \neq j} \frac{\partial}{ \partial \vect{r}_i} \phi_b(r_{jk}
)= \begin{cases}
\frac{1}{2W_{jb}}\sum_{k \neq j} \frac{\partial}{\partial \vect{r}_i} \phi_b(r_{jk}), & \quad \rm{if}\;\;j = i\\
\frac{1}{2W_{jb}}\frac{\partial}{\partial \vect{r}_i}\phi_b(r_{jk}),                             & \quad \rm{if}\;\; j \neq i\;.
\end{cases}
\end{equation}
The pairwise terms are given in Eq.\ref{eq:ppot}, the derivatives are 
straightforward, noting that an atom does not interact with itself so
 $V(r_{ii}) = 0$

%\begin{equation}\frac{\partial}{\partial \vect{r}_i} (U_{pair_{jb}}) =
%\begin{cases}
%\sum_{k \neq j} \frac{\partial}{\partial \vect{r}_i} V_{b} (r_{jk}), & \quad \rm{if}\;\;j = i\\
%\frac{\partial}{\partial \vect{r}_i} V_{b} (r_{jk}), & \quad \rm{if}\;\;j \neq i\;.
%\end{cases}
%\end{equation}
On combining the above equations, the force may then be written:
\begin{equation}
\begin{aligned} 
\vect{f}_i = &- \sum_{j \neq i}  \bigg[ 
 \frac{n_{j1}}{4 W_{j1}}\bigg(\frac{n_{j1}}{N_1} - 1\bigg) \frac{\partial \phi_{1}(r_{ji})}{\partial \vect{r}_i}
 + \frac{n_{j2}}{4  W_{j2}}\bigg(\frac{n_{j2}}{N_2} - 1\bigg) \frac{\partial \phi_{2}(r_{ji})}{\partial \vect{r}_i}\\
&\quad \quad \quad \quad+ n_{j1}  \frac{\partial V_{1}(r_{ji})}{\partial \vect{r}_i}
+ n_{j2}  \frac{\partial V_{2}(r_{ji})}{\partial \vect{r}_i}  \bigg] \\
& - \bigg[ 
 \frac{n_{i1}}{4 W_{i1}}\bigg(\frac{n_{i1}}{N_1} - 1\bigg) \sum_{k \neq i} \frac{\partial \phi_{1}(r_{ik})}{\partial \vect{r}_i}
 + \frac{n_{j2}}{4  W_{j2}}\bigg(\frac{n_{j2}}{N_2} - 1\bigg) \sum_{k \neq i} \frac{\partial \phi_{2}(r_{ik})}{\partial \vect{r}_i}\\
& \quad \quad \quad \quad+ n_{j1} \sum_{k \neq i}\frac{\partial V_{1}(r_{ik})}{\partial \vect{r}_i}
+ n_{j2} \sum_{k \neq i}  \frac{\partial V_{2}(r_{ik})}{\partial \vect{r}_i}
\bigg]\;.
\end{aligned}
\end{equation}
To simplify the appearance of this equation we  define
 \begin{equation}\widetilde{W}_{ib} = \frac{n_{ib}(N_b-n_{ib})}{4N_b W_{ib}}\end{equation}
and the force becomes
\begin{equation}
\begin{aligned} 
\vect{f}_i = & \sum_{j \neq i}  \bigg[ 
\widetilde{W}_{j1} \frac{\partial \phi_{1}(r_{ji})}{\partial \vect{r}_i}
+ \widetilde{W}_{j2} \frac{\partial \phi_{2}(r_{ji})}{\partial \vect{r}_i}
- n_{j1}  \frac{\partial V_{1}(r_{ji})}{\partial \vect{r}_i}
- n_{j2}  \frac{\partial V_{2}(r_{ji})}{\partial \vect{r}_i}  \bigg] \\
& + \sum_{k \neq i} \bigg[ 
\widetilde{W}_{i1} \frac{\partial \phi_{1}(r_{ik})}{\partial \vect{r}_i}
+ \widetilde{W}_{i2} \frac{\partial \phi_{2}(r_{ik})}{\partial \vect{r}_i}
- n_{j1} \frac{\partial V_{1}(r_{ik})}{\partial \vect{r}_i}
- n_{j2}  \frac{\partial V_{2}(r_{ik})}{\partial \vect{r}_i}
\bigg]\;.
\end{aligned}
\end{equation}
The summations in the second set of square brackets may be reindexed in terms of $j$'s and included in the first set of square brackets to give
\begin{equation}
\begin{aligned} 
\vect{f}_i = \sum_{j \neq i}  \bigg[ &
\widetilde{W}_{j1} \frac{\partial \phi_{1}(r_{ji})}{\partial \vect{r}_i}
+ \widetilde{W}_{j2} \frac{\partial \phi_{2}(r_{ji})}{\partial \vect{r}_i}
- n_{j1}  \frac{\partial V_{1}(r_{ji})}{\partial \vect{r}_i}
- n_{j2}  \frac{\partial V_{2}(r_{ji})}{\partial \vect{r}_i}  \\
& +
\widetilde{W}_{i1} \frac{\partial \phi_{1}(r_{ij})}{\partial \vect{r}_i}
+ \widetilde{W}_{i2}  \frac{\partial \phi_{2}(r_{ij})}{\partial \vect{r}_i}
- n_{j1} \frac{\partial V_{1}(r_{ij})}{\partial \vect{r}_i}
- n_{j2}  \frac{\partial V_{2}(r_{ij})}{\partial \vect{r}_i}
\bigg]\;.
\end{aligned}
\end{equation}
Since both $\phi$ and $V$ are pair potentials \( \phi_{b}(r_{ij}) = \phi_{b}(r_{ji})\) and \( V_{b}(r_{ij}) = V_{b}(r_{ji})\). The derivatives of $\phi$ are
\[\frac{\partial}{\partial \vect{r}_i}\phi_b(r_{ij})  = \frac{\partial}{\partial r_i}\phi_b(r_{ij}) \hat{\vect{r}}_{ij} = \phi^{\prime}_b(r_{ij}) \hat{\vect{r}}_{ij}\]
where the  \(\phi^{\prime}_b\) are scalar quantities and \(\phi^{\prime}_b(r_{ij}) = \phi^{\prime}_b(r_{ji})\).  Similarly, the derivatives of $V$ are
\[\frac{\partial}{\partial \vect{r}_i}V_b(r_{ij})  = \frac{\partial}{\partial r_i}V_b(r_{ij}) \hat{\vect{r}}_{ij} = V^{\prime}_b(r_{ij}) \hat{\vect{r}}_{ij}\]
where again the  \(V^{\prime}_b\) are scalar quantities.

Finally, the force on atom i is then \vspace{1 ex}
\newline
\fbox{
\begin{minipage}{\textwidth}
\begin{equation}
%\fbox{$\displaystyle
\begin{aligned}
\vect{f}_i = \sum_{j \neq i}  \bigg[ &
(\widetilde{W}_{i1} + \widetilde{W}_{j1}) \phi^{\prime}_1(r_{ij}) 
+ (\widetilde{W}_{i2} + \widetilde{W}_{j2}) \phi^{\prime}_2(r_{ij}) \\
&- ( n_{i1} +  n_{j1}) V^{\prime}_{1}(r_{ji})
- ( n_{i2} +  n_{j2}) V^{\prime}_{2}(r_{ji})
\bigg]\hat{\vect{r}}_{ij}
\end{aligned}
%$}
\end{equation}
\end{minipage}}
 \vspace{1 ex}

\end{widetext}
\end{document}